\documentstyle[emulateapj,psfig]{article}
\def\n{\kern-.667em}
\def\ltsima{$\; \buildrel < \over \sim \;$}
\def\simlt{\lower.5ex\hbox{\ltsima}}
\def\gtsima{$\; \buildrel > \over \sim \;$}
\def\simgt{\lower.5ex\hbox{\gtsima}}
\def\kms{{\rm\,km\,s^{-1}}}

\def\masyr{{\rm\,mas/yr}}
\def\kpc{{\rm\,kpc}}

\def\msun{{\rm\,M_\odot}}

\newcommand{\fmmm}[1]{\mbox{$#1$}}
\newcommand{\scnd}{\mbox{\fmmm{''}\hskip-0.3em .}}
\newcommand{\scnp}{\mbox{\fmmm{''}}}

\def\s{\ifmmode \widetilde \else \~\fi}
\def\={\overline}

\def\spose#1{\hbox to 0pt{#1\hss}}

\def\etal{{\it et al.\ }}

\def\ie{{ i.e.,\ }}
\def\lta{\mathrel{\spose{\lower 3pt\hbox{$\mathchar"218$}}
     \raise 2.0pt\hbox{$\mathchar"13C$}}}
\def\gta{\mathrel{\spose{\lower 3pt\hbox{$\mathchar"218$}}
     \raise 2.0pt\hbox{$\mathchar"13E$}}}
\def\Dt{\spose{\raise 1.5ex\hbox{\hskip3pt$\mathchar"201$}}}	
\def\dt{\spose{\raise 1.0ex\hbox{\hskip2pt$\mathchar"201$}}}	

\def\=={\equiv}

\def\dotsfill{\leaders\hbox to 1em{\hss.\hss}\hfill}

\def\Gyr{{\rm\,Gyr}}

\slugcomment{Submitted to The Astrophysical Journal Letters}

\lefthead{Ibata \etal}
\righthead{}

\begin{document}

\title{FAINT, MOVING OBJECTS IN THE HUBBLE DEEP FIELD: 
COMPONENTS OF THE DARK HALO?$^1$}

\author{
Rodrigo A. Ibata\altaffilmark{2},
Harvey B. Richer\altaffilmark{3},
Ronald L. Gilliland\altaffilmark{4} \&
Douglas Scott\altaffilmark{3}}

\altaffiltext{1}
{Based on observations with the NASA/ESA Hubble Space Telescope}
\altaffiltext{2}
{European Southern Observatory
Karl Schwarzschild Stra\ss e 2, D-85748 Garching bei M\"unchen, Germany}
\altaffiltext{3}
{Department of Physics \& Astronomy, University of British Columbia,
2219 Main Mall, Vancouver, B.C., V6T 1Z4, Canada.}
\altaffiltext{4}
{Space Telescope Science Institute, 3700 San Martin Drive, Baltimore,
MD 21218 USA.}



\begin{abstract}
The deepest optical image of the sky, the Hubble  Deep Field (HDF), obtained
with the Hubble Space Telescope (HST) in December 1995, has been compared to
a similar image taken in December 1997.  Two  very faint, blue, isolated and
unresolved  objects  are found  to  display  a substantial   apparent proper
motion, 23$\pm$5\,mas/yr and  26$\pm$5\,mas/yr;  a further three objects  at
the detection limit  of  the second epoch observations  may  also be moving.
Galactic structure  models   predict a general   absence  of  stars in   the
color-magnitude  range  in which these  objects  are  found.  However, these
observations  are consistent with   recently-developed models  of old  white
dwarfs with  hydrogen  atmospheres,   whose  color, contrary    to  previous
expectations, has been shown to be blue.  If these apparently moving objects
are  indeed old  white  dwarfs with   hydrogen  atmospheres and masses  near
0.5\,M$_\odot$,  they have ages of  approximately  12\,Gyr, and a local mass
density that is sufficient, within the large  uncertainties arising from the
small   size of the   sample,  to account   for  the entire missing Galactic
dynamical mass.
\end{abstract}

\keywords{Galaxy: halo -- solar neighbourhood -- dark matter -- stars: white dwarfs}

\section{Introduction}

The MACHO  collaboration have found $\sim 16$   massive compact halo objects
(MACHOs)   in  their 4  year  survey   for  microlensing  towards  the Large
Magellanic Cloud (Alcock  \etal\  1997).  Based   on the standard   multiple
component mass model for  the Galaxy (Griest 1991),  they conclude that this
lensing  rate  is much higher  than   the Galactic  thick  disk and Galactic
stellar halo can  account for, and   argue that the  lenses are stellar-mass
objects that reside in an extended  halo around the  Galaxy and constitute a
sizeable  fraction, perhaps half, of  its mass.  This population should also
be present in the Solar neighborhood, and in principle could be seen in deep
or wide  field   surveys,  if  the  individual components  are  sufficiently
luminous.

We set out  to investigate whether the  HDF (Williams \etal\ 1995; hereafter
W95) could be used, in conjunction  with images in the  same field taken two
years later, to measure  proper motions (PMs) of  faint stars that could  be
nearby MACHO   candidates.  Though the  HDF  covers  a  small  area, $0.0015
\Box^\circ$, it is very deep (reaching to ${\rm V  \sim 28}$), so the volume
probed  for faint  Galactic  sources  is  considerable.   Earlier  starcount
studies  in the HDF  (Flynn  \etal\ 1996;  Elson  \etal\ 1996; M\'endez \etal\
1996)  were not  able to  reach to  the faint limit  of  the  dataset due to
star-galaxy confusion at magnitudes  fainter   than $\rm  V  =  26.5$.   PMs
provide  a means   to perform   further  star-galaxy discrimination,   since
anything with significant PM cannot be very distant.

\section{Registration of the Hubble Deep Field frames}

Our second epoch HDF exposures, obtained in December 1997  in the same field
as  the original HDF, give a  baseline of almost exactly  two years and have
total integration of 27.3~Ksec in U (F300W) and  63~Ksec in I (F814W). These
exposures were   obtained  at  several  slightly  offset positions,  and  at
approximately the same orientation as the originals.   For the present work,
we used only the I-band data to determine PMs.

The PMs of stars are measured as an angular displacement between epochs with
respect to some reference frame; the accuracy of this frame is therefore one
of   the    crucial limiting factors  affecting    the  accuracy  of  the PM
measurements.   The  determination of   an accurate reference  frame is  not
entirely  straightforward due to the optical  distortions of the cameras and
irregularities in   the construction of  the  detectors.  Fortunately, these
instrumental signatures are stable, and   so their  effects can be   largely
eliminated.  The  approach we took  for constructing the  reference frame in
each field was to obtain differential measurements of the positions of $\sim
50$ bright, compact galaxies on each frame, and, by adopting a model for the
optical distortion of  the WFPC2 mosaic  (Trauger \etal\ 1995),  determine a
simple linear transformation relating the  distortion-corrected frames.  The
accuracy   of this registration  is   better than  2 milli-arcseconds (mas),
judging from the rms scatter in the positions of the reference galaxies.  In
order to   obtain an independent  check of  the   registration precision, we
created a set of new  frames by resampling  the original data onto the  same
grid as  the reference image adopted  by  W95, using the  optical distortion
model and the  computed offsets.   Cross-correlation  of the images of  both
epochs against the reference image showed them to be  aligned to an accuracy
of better than 3~mas.

\section{Proper motion measurements}

A  maximum likelihood image center finding  algorithm was developed for this
project (Ibata \& Lewis 1998). The important  advantage of this technique is
that it  does not use  a combined frame  for calculating the image centroids
(avoiding    the    inevitable    degradation   of    information)     or  a
resolution-enhanced   combined  frame (which   avoids also having correlated
noise).  By  applying  this  technique to the   HDF data,  they showed  that
centroiding  ($1\sigma$) accuracies of  $\sim  10$~mas  can be obtained  for
stars of magnitude ${\rm I \sim 28}$, degrading to  $\sim 1$~mas for ${\rm I
< 25}$.

An input list of 443 compact  objects (detected with DAOPHOT, Stetson 1987),
was  provided    to  the  PM  algorithm.   In   addition   to   finding  the
maximum-likelihood image centroids, the routine  returns both the likelihood
surface and $P(  R < 2.5 )$, the  probability that the  image data (within a
radius of $2.5$~pixels from the most likely centroid position) is drawn from
the same distribution as the PSF. Objects  for which $P( R  < 2.5 ) < 0.01$,
were considered to be extended and therefore discarded.

In this way, we detected 58 isolated point-sources, of which 40 have ${\rm V
> 27}$.  The brighter (${\rm  V < 27}$)  moving sources can be accounted for
by  known Galactic stellar populations.  The  calibrated V, ${\rm  (V - I)}$
color-magnitude diagram (on the AB  system) of the  faint subset is shown in
Figure~1.  The PM vectors are also  plotted, either as black
or green arrows.  The axes have been chosen such that motion in the direction
of Galactic longitude is parallel to the (${\rm V - I}$) color axis, whereas
motion in  the direction of Galactic  latitude is parallel  to the magnitude
axis (increasing downwards).  The PM  scale on this   plot is such that  one
abscissa or ordinate unit represents a motion  of 100~mas (approximately one
pixel on the Wide Field chips) over the two year baseline of the experiment.
The red arrows are of length equal to the PM uncertainty in the direction of
the PM.

Among the population of 40 faint point sources with ${\rm V > 27}$, five are
observed to have  PMs that  exceed the measurement uncertainty by
more than a factor  of 3.  These objects are  listed in Table~1.  On the top
row of Figure~2 we display co-added images, $4\scnp$ on a side, of the field
around each of these objects; the middle two rows are high-resolution images
which clearly show the centroid shifts between epochs;  the bottom row shows
the centroid likelihood contours given the data at each epoch.

\section{Checks of the centroid shifts}

Simulations were carried out to constrain the incompleteness of the observed
sample and to check the PM uncertainties.  In each  simulation, we added 100
fake stars, of  zero PM, into all the   individual V and I-band  frames. The
stars were added uniformly between ${\rm I = 25}$ and ${\rm I = 31}$, with a
color of ${\rm  V - I = 0}$.   As before, we  calibrated the photometry, and
rejected those objects for  which $P( R <   2.5 ) <   0.01$, those that  had
photometric uncertainties $\delta({\rm V}) > 0.5$,  $\delta({\rm I}) > 0.5$,
and those with a  neighbor less than 1 arcsec  away. Twenty such simulations
were performed on   each  of the  Wide  Field chips   for a total   of  6000
artificial  stars.   The completeness of our  sample  at  magnitudes between
${\rm I = 27}$ and ${\rm I = 28}$ is $42\% \pm 2\%$.  It  was found that the
maximum deviation between epochs in the computed centroids of the artificial
stars was  $18\masyr$,   with the biggest deviation  being   $3.6$ times its
estimated uncertainty. The PMs of our five objects exceed even these maximum
random values,  indicating that the measured  displacements are  unlikely to
occur by chance.

But what about systematic effects on our PM measurements?  There are several
possible mechanisms  which could have  given rise to  a false  PM detection.
The most obvious are cosmic rays,  hot pixels, the inevitable Poisson noise,
and construction irregularities of the detectors.  The artificial star tests
provide a convenient  way to test these  concerns.  The simulated stars were
placed into the individual data frames at exactly the  same location at both
epochs,  so if by  chance that location corresponds  to  a position where an
unmasked contaminant  remains, a false PM  could be detected.  We  find that
there  is less  than a  0.1\% chance for  displacements of  the magnitude of
those detected in the five faint objects listed  in Table~1, so it is highly
unlikely that the measured PMs have been affected in this way.

Nevertheless, we ran a further  check to make  sure that the PM measurements
of the  five apparently  moving objects  are not  influenced by large  noise
spikes or defective pixels.  In  each test, we rejected  all the data at one
spacecraft pointing (i.e., `dither'), and re-computed the maximum-likelihood
centroids. If there  was a residual unmasked  cosmic ray or  a hot pixel, or
even a particularly high Poisson deviation that was  causing an incorrect PM
measurement, the contaminant would be absent in one of the test runs, and so
we would find a PM much closer to zero in that test run.  However, the tests
did not  reveal such  an effect,  confirming  that our  results are   not an
artifact of residual cosmic rays, hot pixels or Poisson noise.
 
To  check the five apparently high  PM detections, we have implemented three
other, more direct, techniques that operate  on combined images; the results
of these tests are listed in Table~1.   We find that  the PM measurements of
objects 4--551, 2--766, and 4--492 are reproduced, within the uncertainties,
by all four techniques.  However, the PMs of 4--141 and 2--455 measured with
the  maximum-likelihood method are not verified  by  the other methods.  For
this  reason we do  not consider  the PM  measurements  of the faint objects
4--141 and 2--455 to be secure.   Nevertheless, we include  all 5 objects in
our   discussion, since they   represent   the  full  sample of   candidates
discovered in our systematic survey for PMs in the HDF.

\section{The nature of the faint, apparently moving, sources}

A natural explanation for the apparent large PMs of the five objects is that
they could  be   detections of  supernova (SN)    events  in high   redshift
star-forming galaxies.  If the   SNe were somewhat off-centered  from  their
host galaxy, this would skew the light distribution  at one epoch, causing a
shift in the computed image centroid  that is not  due to actual PM.  Indeed
there is weak evidence that object 2--766 became fainter, by $25\% \pm 13\%$
between  the two  epochs, which could  be  explained within this paradigm as
being due  to a SN   in the first  epoch   dataset.  However, there  was  no
significant    brightness variation in   objects  4--551, 4--141, 4--492 and
2--455, implying that if there were SNe present in one of the epochs in each
of these three  hypothetical galaxies, they must  have been extremely faint.
Nevertheless,  we   should   ask:   is   it possible  that     the  computed
maximum-likelihood  object   centroids could   have   been  so substantially
affected by such faint SNe to have given rise to the observed large apparent
positional shift?

To answer  this we conducted additional  artificial star tests, adding point
sources within $0\scnd2$ of the moving candidates  to simulate the effect of
SNe on    the  measurements of   object   centroids.  New maximum-likelihood
centroid  positions were  calculated from   these modified  frames and  this
experiment was repeated 1000 times for each of the objects.  We find that it
is highly improbable  (with  a chance  of less  than   1 in 1000)  that  the
centroid shifts observed in objects  4--551, 4--141, 4--492 and 2--455 could
be due to SNe offset from their host galaxy  centers, as the brightnesses of
these  hypothetical events, constrained   by  the small  observed brightness
variations between epochs, are too small to affect the image profiles to the
required extent.  However, we find  that there is  a  $5$\% chance that  the
apparent motion of object 2--766 can be explained by the SN hypothesis.

Thus, with the  caveat that there are no  pathological  frame distortions on
the scale of 1 to  2 arcsec, we are  forced  to conclude that object  4--551
shows significant PM, so it must be a  nearby moving source, and that object
2--766 likely has  measurable PM, though there  is  a small chance  that the
measured offset   is an artefact of a   SN superimposed on   a  galaxy.  The
faintest three candidates are low  signal-to-noise photometric detections in
the second epoch frames, so their PM  measurements are more likely to suffer
from unknown systematic errors.  With these further caveats, it appears that
object 4--492  has  appreciable PM, while   objects 4--141  and  2--455 have
significant   motion using what   we   consider to  be  the  most  sensitive
procedure.  For these sources, the  detected PMs are  much too low for Solar
System objects, and  since the two epochs  were at almost identical times of
the year, parallax  is also  ruled out.   The only plausible  alternative is
that at least some  of this sample of five  objects are Galactic stars. This
conclusion is supported by M\'endez  \& Minniti (1999),  who find that faint
blue point sources are approximately twice as numerous in the HDF South than
in the HDF (the HDF-S  is located  $56^\circ$ from  the galactic center,  as
opposed to $110^\circ$ for the HDF).

\section{Faint, moving stars?}

What  type of star  could these sources  be?  If they  were Galactic disk or
thick  disk stars they would  have to be intrinsically faint  in order to be
seen  at ${\rm V  \sim 28}$, since any  intrinsically bright  stars would be
several tens of kiloparsecs out of the plane of the  disk, and therefore not
disk members.   Indeed, stars with  vertical heights of $z  < 10 \kpc$ above
the plane of the Galaxy must have absolute magnitudes $M_{\rm  V} > 12.6$ to
be observed  at ${\rm V  =  28}$.  All known main-sequence  populations  are
extremely red in ${\rm V - I}$ at these magnitudes.  Disk white dwarfs (WDs)
at the observed ${\rm V - I}$ colors would be about 1 Gyr old and have $M_V$
near  13.  This would  again  place them at  about  10 kpc above  the plane.
However,  our Galactic structure  model (Ibata 1995)  predicts an absence of
disk, thick disk  (both including WDs), or local  spheroid stars blueward of
${\rm V-I =  1.0}$   at ${\rm  V \sim  28}$.   Luminous spheroid  stars   at
distances of $\sim 100 \kpc$ may be found in the color-magnitude range ${\rm
27 < V < 29}$, ${\rm V - I < 1.0}$, but our  Galaxy model predicts only 0.01
such stars in the  HDF (though we caution the  reader that no model has been
properly  tested in this regime). However,  due to their enormous distances,
the PMs of such stars would not have been detectable in our experiment.  The
observed moving stars  therefore cannot belong to  known disk, thick disk or
spheroid populations.

\section{Ancient white dwarfs of the Galactic halo?}

No halo WDs were detected in the most sensitive  survey to date (Knox \etal\
1999), though it is possible that their detection limit was  not as faint as
claimed. So   can   the  moving objects   we  have   detected  be halo   WDs
nevertheless?   That  the  halo may contain   numerous  such stars  has been
suggested naturally (Kawaler 1996;  Chabrier 1999) through the  microlensing
experiments    which  yield  MACHO   masses    (Alcock   \etal\  1997)    of
$0.5^{+0.3}_{-0.2} \msun$, a  value similar to  the mass  of $0.51 \pm  0.03
\msun$ inferred for old WDs in ancient star clusters (Richer \etal\ 1997).

Until recently, cooling models of old WDs predicted  that these stars should
be  quite red.  However, new theoretical  work (Hansen 1998, 1999; Saumon \&
Jacobson 1999), which  extends the  effective  temperatures of WDs  to below
$4000\,$K,  indicates that  H$_2$ provides  strong opacity  in the infrared,
forcing the radiation out in the blue.  Very old WDs, of  age $12 \Gyr$, and
mass $0.5 \msun$, have ${\rm V - I \sim 0.2}$ according to these models.
 
If the entire dark matter halo  of the Milky Way were  to be made up of such
WDs there should be approximately 9 such objects in the HDF between ${\rm 27
< V < 28.5}$,  they should have  colors ${\rm -0.2  < V-I < 1.0}$, and  they
should be situated at a mean distance  of $1.2 \kpc$.  Assuming an intrinsic
one-dimensional  velocity dispersion  of    $200 \kms$ for  the   dark  halo
population  in the  Solar Neighborhood, with   zero  net rotation  about the
Galactic center, the expected  PM  distributions (after correction  for  the
Solar Reflex  Motion) in both $\mu_\ell$  and $\mu_b$, have  a mean  of $-20
\masyr$ and a dispersion of $35 \masyr$.

The sample of  moving objects we have  discovered fits reasonably  well into
this  model:  correcting for the  $42\%  \pm  2\%$  completeness of  the HDF
dataset between ${\rm 27 < I < 28}$, a total of  about 4 stars are expected;
their colors  and magnitudes also  agree with this model;  and their PMs are
consistent  (at the  20\%  level)  with   being   drawn from  the   expected
distributions.  Two of them (4-551  and 4-492) also  appear to have spectral
energy distributions consistent (within the large photometric uncertainties)
with them being old WDs (see Table~1).

This suggests  that  we may have discovered,  through  their apparent PMs, a
population of ancient WDs that are the local counterparts of the MACHOs.  If
this conclusion is  correct,  a  substantial fraction   of the dark   matter
concentrated in the inner  regions of galactic halos  would be baryonic  and
locked up in the form of very faint, blue (and therefore, ancient) WD stars.
Further work is required to ascertain whether the  problems of this scenario
can   be overcome,  notably   the requirement  that  stars   form in  a very
restrictive mass  range in the  early Universe (Tamanaha \etal\  1990), that
the   precursors of the  WDs would  make young   galaxies appear anomalously
bright (Charlot \& Silk 1995), and that the interstellar medium would become
over-enriched with  heavy elements due to the  material  ejected from the WD
progenitors (Gibson \&  Mould 1997).

By regarding   our PM measurements as predictions,   a direct test   will be
possible  with  third epoch   observations of  the  HDF,  to be  obtained in
December 1999 with the HST.  Ultimately, however, spectroscopic observations
will be needed to  show whether or  not these objects  are WDs; this may  be
feasible if nearer members of this population can be discovered.

\begin{acknowledgements}
The research of HBR and DS is supported in part by  the Natural Sciences and
Engineering   Research Council of  Canada.   RLG  is  supported under  grant
GO-6473.01-95A from STScI.
\end{acknowledgements}

\begin{figure*}
\psfig{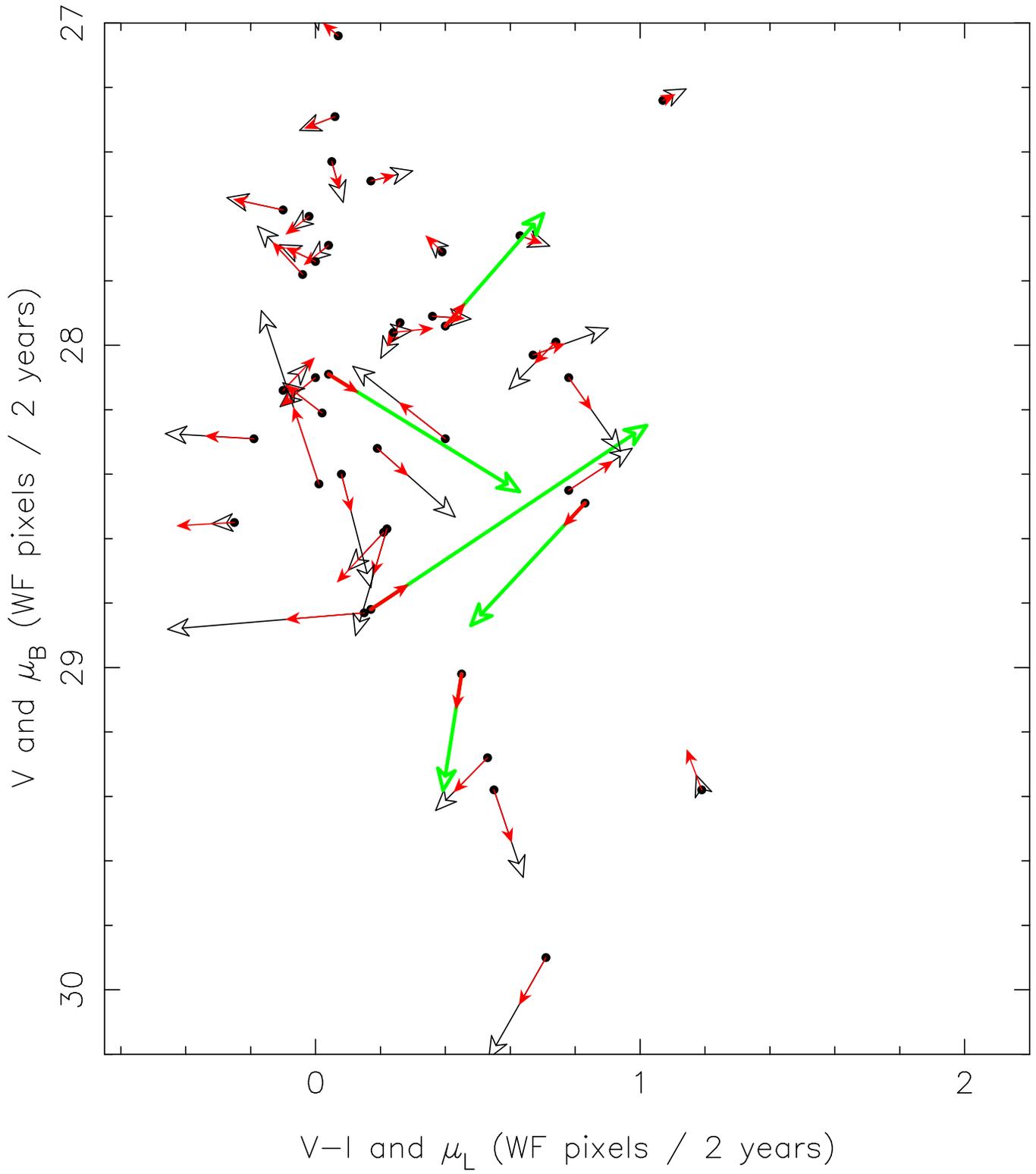}
\caption{The faint end of the color-magnitude diagram of the HDF showing the
40  faint unresolved  objects together with their PMs.}
\end{figure*}

\begin{figure*}
\psfig{figure=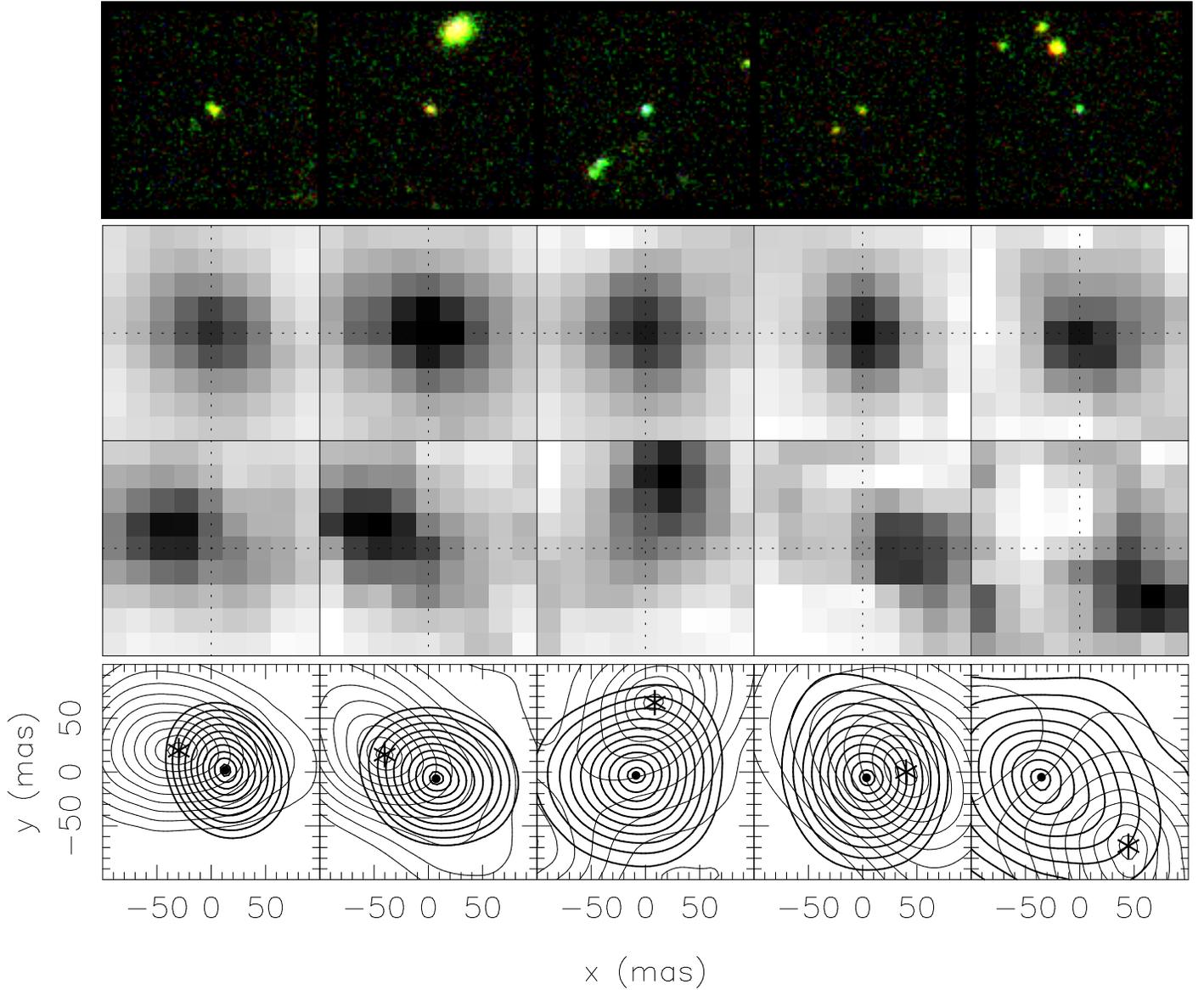,bbllx=45pt,bblly=330pt,bburx=275pt,bbury=531pt,width=15cm,angle=0}
\caption{ The five faint, apparently moving objects.   The panels show, from
left to right, images and likelihood contours of the five faint objects with
significant PM: 4--551, 2-766, 4--141, 4--492 and  2--455.  The color images
in the top row show the immediate field ($4\scnp \times 4\scnp$) around each
object; RGB  intensities indicate,  respectively, the  fluxes in  the F814W,
F606W and F450W   filters.  High  resolution  images of   these objects were
constructed by  using  the  PSF  model  to redistrubute  the flux  from  the
individual WF frames onto a $4\times$ oversampled grid (\ie\ $0\scnd025/{\rm
pixel}$);  these images,   for the  first   and second  epoch  datasets  are
displayed, respectively, in  the second and  third  rows. The same  $0\scnd2
\times0\scnd2$ region of  the sky is shown  for each object, and cross-hairs
have been  added at the position corresponding  to the centroid  location in
the first epoch  image.  The bottom row  shows the  corresponding likelihood
contours of  the object centroids obtained from  the first epoch data (thick
lines) and the  second epoch data (thin lines).   The contour intervals  are
such  that the $n$th  contour  marks the boundary  of  the region where  the
likelihood has fallen by a factor of $\exp{-{{n^2} \over  2}}$ from the most
likely value.}
\end{figure*}

\begin{deluxetable}{cccccccllrrrr}
\tablefontsize\footnotesize
\tablenum{1}
\tablecaption{Faint stars with detected  proper motion.
}
\tablewidth{0pt}
\tablehead{
\colhead{ID} & 
\colhead{I} & 
\colhead{V-I} & 
\colhead{B-V} & 
\colhead{U-B} & 
\colhead{V-J} &
\colhead{$\mu_\ell,\mu_b$} &  
\colhead{$d$} & 
\colhead{$v$} &
\colhead{Method 1} &
\colhead{Method 2} &
\colhead{Method 3} &
\colhead{Method 4} \nl
 & & & & & & & & &
\colhead{$\delta x$,$\delta y$} &
\colhead{$\delta x$,$\delta y$} &
\colhead{$\delta x$,$\delta y$} &
\colhead{$\delta x$,$\delta y$} \nl
 & & & & & &
\colhead{($\masyr$)} & 
\colhead{(kpc)} & 
\colhead{(km/s)} &
\colhead{($\masyr$)} &
\colhead{($\masyr$)} &
\colhead{($\masyr$)} &
\colhead{($\masyr$)} }
\startdata
\n 4--551 &\n27.54&\n0.40&\n$>$1.76&\n\dots  &\n-0.98
 &\n~15.1(4.5),-17.4(5.0)&\n1.3(0.1)&\n 220(60)
 &\n\n\n-21(5),~~9(5)&\n\n-14(9),~5(9)&\n\n-19(2),~4(2)&\n\n-20(9),~~8(9)\n\nl
\n 2--766 &\n27.66&\n0.83&\n0.18   &\n-0.09  &\n\dots
 &\n-17.6(4.9),~19.0(5.1)&\n1.7(0.3)&\n 280(100)
 &\n\n\n-24(5),~11(5)&\n\n-21(9),~1(9)&\n\n-15(5),~4(4)&\n\n-21(9),~~3(9)\n\nl
\n 4--141 &\n28.05&\n0.04&\n-0.24  &\n0.47   &\n\dots
 &\n~29.4(5.3),~18.2(6.0)&\n1.4(0.2)&\n 400(80)
 &\n\n\n~~9(6),~34(5)&\n\n~-1(9),16(9)&\n\n~~3(5),~9(4)&\n\n~~5(9),~18(9)\n\nl
\n 4--492 &\n28.57&\n0.45&\n$>$1.07&\n\dots  &\n-0.17
 &\n~-2.9(6.5),~18.1(5.6)&\n2.2(0.4)&\n 320(120)
 &\n\n\n~18(6),~~3(6)&\n\n~17(9),-6(9)&\n\n 20(5),~6(4)&\n\n21(9),~~8(9)\n\nl
\n 2--455 &\n28.65&\n0.17&\n0.37   &\n$>$0.00&\n\dots
 &\n~42.5(7.4),-28.6(6.3)&\n2.0(0.3)&\n 520(160)
 &\n\n\n~40(6),-32(8)&\n\n-13(9),-8(9)&\n\n~2(15),4(11)&\n\n~-1(9),-12(9)\n\nl
\n model  &\n\dots&\n0.38&\n1.51   &\n0.70   &\n-0.38
 &\n   \dots,\dots     &\n\dots   &\n\dots    
 &\n\dots,\dots & \n\dots,\dots& \n\dots,\dots& \n\dots,\dots \n\nl
\enddata
\tablecomments{Column (1)  notes  the W95 identification  label, and columns
(2) to (6) give the calibrated isophotal AB photometry from W95 and Thompson
{\it et al.} (1999).  Column (7) lists the PM  results, where $\mu_\ell$ and
$\mu_b$ are  the PMs in the direction   of, respectively, Galactic longitude
and Galactic latitude.  The last row in  the table labelled `model' contains
the expected colors from a $0.5\,{\rm  M}_{\odot}$ hydrogen-rich white dwarf
with an effective temperature of 3000\,K (Hansen 1999).  Columns (8) and (9)
list,  respectively, the  distance  and velocity   (corrected  for the Solar
reflex motion) inferred from that  white dwarf model.   Columns (10) to (13)
show the PM results   from different  measurement techniques.   The  $\delta
x$,$\delta y$ offsets listed are parallel to the  CCD axes.  Method 1 is the
maximum-likelihood technique  applied to individual  frames. Method 2 is the
positional difference in a Gaussian PSF model  fit to the object profiles in
the  combined frames   at  each  epoch.   Method   3 is   a  two-dimensional
cross-correlation  of a small $3\scnp \times  3\scnp$ region of the combined
images  at  the two  epochs  around  each of  the  objects  (with a $1\scnp$
cosine-bell  tapering region applied  to   the edges).   Finally, method   4
involves  a Gaussian  PSF fit  to the object   profiles, but this time using
images combined with four-times oversampling in  an independent analysis for
a separate project (Gilliland \etal\ 1999).  The  uncertainties in methods 2
and 4 have  been estimated by  comparing two  equal exposure subsets  of the
first epoch  dataset, where there is   no possibility of  real  motion.  The
uncertainty in method 3 is derived from a fit to the correlation function.}
\end{deluxetable}

\end{document}